\documentclass[aps,pra,reprint,groupedaddress,showpacs]{revtex4-1}
\usepackage{eurosym}
\usepackage{amsmath}
\usepackage{graphicx}
\usepackage{textcomp}
\usepackage{hyperref}
\usepackage{amsfonts}
\usepackage{amssymb}

\begin{document}

\title{Energy backflow in unidirectional spatiotemporally localized wavepackets}
\author{Ioannis Besieris$^{1}$}
\author{Peeter Saari$^{2,3}$}
\affiliation{$^{1}$The Bradley Department of Electrical and Computer Engineering,
Virginia Polytechnic Institute and State University, Blacksburg, Virginia
24060, USA}
\affiliation{$^{2}$Institute of Physics, University of Tartu, W. Ostwaldi 1, 50411,
Tartu, Estonia}
\affiliation{$^{3}$Estonian Academy of Sciences, Kohtu 6, 10130 Tallinn, Estonia}
\email{Corresponding author: peeter.saari@ut.ee}
\date{\today }

\begin{abstract}
Backflow, or retro-propagation, is a counterintuitive phenomenon where for a
forward-propagating wave the energy locally
propagates backward. In this study the energy backflow has been examined in connection
with relatively simple causal unidirectional finite-energy solutions of the wave equation 
which are derived from a factorization of the so-called basic splash mode. 
Specific results are given for the
energy backflow arising in known azimuthally symmetric unidirectional wavepackets, 
as well as in novel azimuthally asymmetric extensions. Using the Bateman-Whittaker
technique, a novel finite-energy unidirectional null localized wave has been
constructed that is devoid of energy backflow and has some of the topological
properties of the basic Hopfion.
\end{abstract}


\maketitle

\section{Introduction}

In general terms, the phenomenon of backflow takes place when some quantity
(probability or energy density flow, local momentum, etc.) in some
spatio-temporal region of a wavefield is directed backward with respect to
the directions of all plane-wave constituents of the wavefield \cite{Berry,
IBB}. Position probability backflow specific to quantum particles,
such as electrons, has been termed `quantum backflow,' and this subject is
actively studied (see the newest review \cite{Bracken} and references
therein). A dispute arose recently over the distinction between the quantum
backflow and backflow phenomena known in classical field theories \cite{IBB,
Comment}. In our opinion, the contradistinction here is largely of a
terminological nature. At least nobody doubts that backflow is a wave
phenomenon that may occur in all kinds of wavefields, particularly in those
describable by the Schr\"{o}dinger or the Maxwell or the wave equations in
free space. 

Indeed, already a simple field of four appropriately polarized and directed
electromagnetic plane waves exhibit prominent energy backflow 
described by the Poynting vector whose direction is reversed with respect to 
the direction of propagation of the resultant wave \cite{Katz,quartet,meiequartet}. 
In the physical optics community the energy
backflow in sharply focused light has been known more than an half a century
and has been thoroughly studied theoretically recently \cite{negSfocal,
negSfocaluus,negSfocaluus2}. In the context of quantum backflow,
monochromatic optical fields have been used for recent experimental
verification of the effect \cite{BFexp1,BFexp2}. Energy backflow in
electromagnetic Bessel beams has been analytically demonstrated in \cite{AriBBnegS,TM+TE}. 
In the context of our present subject, important is the theoretical study \cite{TM+TExwave} 
of backflow in pulsed electromagnetic X waves, which belong to the class of the so-called 
localized waves.

Localized waves (LWs)---also known as space-time wavepackets
(STWP)---have been studied intensively during the past thirty years (see
\cite{LW1,LW2,LW3,LW4,LW5,LW6,LW7,LW8,LW9,LW10,LW11,LW12,LW13,LW14,LW15} for pertinent 
literature). They constitute spatiotemporally localized
solutions to various hyperbolic equations governing acoustic,
electromagnetic and quantum wave phenomena and can be classified according to
their group velocity as luminal LWs, or focus wave modes (FWM), superluminal
LWs, or X waves (XWs), and subluminal\ LWs. Further details can be found in
two edited monographs on the subject \cite{LW16,LW17} and in the recent thorough
review article \cite{LW18}. In general, both linear and nonlinear LW pulses exhibit
distinct advantages in comparison to conventional quasi-monochromatic
signals. Their spatiotemporal confinement and extended field depths render
them especially useful in diverse physical applications. Experimental
demonstrations have been performed in the acoustical and optical regimes \cite{LW5,
LW7,LW10,LW19,LW20,LW21,LW22,LW23}. Work, however, has been carried out at microwave 
frequencies recently \cite{LW24,LW25,LW26}.

An important question---widely discussed especially in the early stages of the
theoretical study of LWs at the end of the last century---has been the
physical realizability of localized waves. For example, two- or
three-dimensional electromagnetic luminal localized waves in free space involve the
characteristic variables $\varsigma =z-ct$ and $\eta =z+ct$ of the
one-dimensional scalar wave equation. Consequently, they contain both
forward and backward propagating components. Tweaking free parameters
appearing in the wavepackets can significantly reduce the backward
components. However, it is principally crucial whether the plane-wave
constituents of the wavepacket propagate only in the positive $z$-direction
or also backward. In the first case not only the wave can be launched from an
aperture as a freely propagating beam but, also, the very question of energy
backflow is meaningful.

In the literature, the LWs with forward-propagating plane-wave constituents
have been called somewhat misleadingly 'causal'. In the following discussion we 
shall use the term 'unidirectional'. It must be pointed out, however, that, in
general, a wavepacket as a whole can propagate in the positive $z$-direction
despite the fact that its plane-wave constituents are omnidirectional; and 
\textit{vice versa}: the group velocity of a packet may have a negative $z$-component
despite the fact that the $z$-components of the wavevectors of all its plane-wave 
constituents are positive. Also, for LWs the group velocity typically differs from 
the energy velocity, see \cite{MinuPRA2018,POY2019}.

Our aim in this article is to theoretically study several representatives of a
class of causal, purely unidirectional  \textit{finite-energy} localized waves with
particular emphasis on their energy backflow characteristics. Specifically,
we shall try to ascertain the role of the vector nature and polarization
properties of a light field in the emergence of the backflow effect and its
strength. 
The paper is organized as follows. In the next section we consider several
finite-energy unidirectional localized waves known from the literature.
Extended unidirectional LWs will be introduced in Sec. 3. Sec. 4 is devoted to
a detailed analysis of the backflow characteristics of a unidirectional
vectorial LW derived from the so-called splash mode solution of the scalar
wave equation. This LW is a generalization of the unidirectional 
solution used by Bialynicki-Birula \textit{et al.} \cite{IBB} to
demonstrate the existence of the backflow in electromagnetic fields. Section 5 
is devoted to the derivation of a finite-energy Hopfion-like spatiotemporally
localized wave that is devoid of energy backflow. Concluding remarks are made
in Sec. 6.

\section{Causal, Scalar Unidirectional Localized Waves}

The feasibility of a finite-energy, causal, unidirectional localized wave
was first addressed by Lekner \cite{LW27} using the Fourier synthesis
\begin{widetext}
\begin{equation}
\begin{split}
\psi _{+}\left( {\rho ,z,t}\right) &=\int_{0}^{\infty }d{k_{z}{e^{i{k_{z}}z}}
\int_{0}^{\infty }{dk\,{e^{-ikct\;}}}\int_{0}^{\infty }{d\kappa \,\kappa }
 {\,J_{0}}\left( {\kappa \rho }\right) \delta \left( {-{\kappa ^{2}}
-k_{z}^{2}+{k^{2}}}\right) F\left( {\kappa ,{k_{z}},k}\right) }\\
&={\int_{0}^{\infty }{dk}\,{e^{-ikct\;}}\int_{0}^{k}{d{k_{z}}}{e^{i{k_{z}}z}}
 \sqrt{{k^{2}}-k_{z}^{2}}{\,J_{0}}\left( {\rho \sqrt{{k^{2}}-k_{z}^{2}}}
\right) {F_{1}}\left( {{k_{z}},k}\right) .}  \label{1}
\end{split}
\end{equation}
\end{widetext}
Choosing the spectrum 
\begin{equation}
{F_{1}}\left( {{k_{z}},k}\right) ={\frac{{{k_{z}}{e^{-ka}}}}{\sqrt{{k^{2}}
-k_{z}^{2}}}},\; a>0,  
\end{equation}
one obtains 
\begin{multline}
{\psi _{+}}\left( {\rho ,z,t}\right) =\\\int_{0}^{\infty }{dk}{e^{-k(a+ict)}}
\,\int_{0}^{k}{d{k_{z}}}{k_{z}}{e^{i{k_{z}}z}}{J_{0}}\left( {\rho \sqrt{{%
k^{2}}-k_{z}^{2}}}\right) .  \label{3}
\end{multline}
Carrying out the integrations, Lekner has derived the causal unidirectional solution 
\begin{multline}
\psi_{+}\left(  \rho,z,t\right)  =\\\\{\frac{a^{4}\widehat{a}}{3}\frac{3\left(
\widehat{a}^{2}+\rho^{2}\right)  ^{2}-6z^{2}\left(  \widehat{a}^{2}+\rho
^{2}\right)  -z^{4}+8iz\left(  \widehat{a}^{2}+\rho^{2}\right)  ^{3/2}%
}{\left(  \widehat{a}^{2}+\rho^{2}\right)  ^{3/2}\left(  \widehat{a}^{2}%
+\rho^{2}+z^{2}\right)  ^{3}}},  \label{4}
\end{multline}
with $\tilde{a}=a+ict.$ By interchanging the order of integrations in Eq.~(3)
and using  table integrals from \cite{LW29}, p.~191, No.~10 and p.~133, No.~3, 
we obtained a more compact expression for the same unidirectional solution: 
\begin{equation}
\psi\left(  \rho,z,t\right)  =\frac{a^{4}\widehat{a}\left(  3\sqrt{\rho
^{2}+\widehat{a}^{2}}-iz\right)  }{3\left(  \rho^{2}+\widehat{a}^{2}\right)
^{3/2}\left(  \sqrt{\rho^{2}+\widehat{a}^{2}}-iz\right) ^{3}} \:. 
\end{equation}%
A formal study of a causal, unidirectional localized wave was undertaken by
So, Plachenov and Kiselev \cite{LW28} starting from the finite-energy luminal
splash mode \cite{LW3,LW4} 
\begin{equation}
\psi \left( {\rho ,z,t}\right) =\frac{1}{{\left[ {{a_{1}}+i\left( {z-ct}
\right) }\right] \left[ {{a_{2}}-i\left( {z+ct}\right) }\right] +{\rho ^{2}}}}\;,
\end{equation}
where ${a_{1,2}}$ are positive free parameters. As mentioned in the
introduction, this is a bidirectional solution. However, for ${a_{2}}>>{a_{1}%
}$, the backward components are significantly reduced. So \textit{et al}. \cite{LW28}
decomposed the denominator in Eq. (6) as follows: 
\begin{multline}
\psi \left( \rho ,z,t\right) =\\=\frac{1}{z_\ast ^{2}-S^{2}}
=\frac{1}{S}\left( \frac{1}{z_\ast -S}-\frac{1}{z_\ast +S}
\right) =\frac{1}{2}\left( u_{+}-u_{-}\right) ;\\
S=\sqrt{ct_\ast^{2}-\rho ^{2}};\;\\t_{\ast }=t+\frac{i\left( 
a_{2}+a_{1}\right) }{2c},\;\;z_\ast =z+\frac{i\left( a_{2}-
a_{1}\right) }{2}. 
\end{multline}%
They proved that the expression ${{u}_{+}}={{S}^{-1}}{{\left( {{z}_{*}}-S \right)}^{-1}}$ 
satisfies the wave equation and is solution propagating in the positive $z-$direction. 
On the other hand, ${{u}_{-}}$is a solution propagating purely in the negative $z-$direction. 
They did not analyze whether ${{u}_{+}}$ is unidirectional (causal) in the sense 
we use here, that all its plane-wave constituents propagate solely into the hemisphere 
with ${{k}_{z}}>0$. 

Independently, Bialynicki-Birula \textit{et al}. \cite{IBB} introduced
two luminal unidirectional wavepackets 
\begin{equation}
{h_{\pm }}={\frac{1}{2}}{\frac{1}{\sqrt{{c^{2}}{{\left( {a+it}\right) }^{2}}+%
{\rho ^{2}}}}}{\frac{1}{{\sqrt{{c^{2}}{{\left( {a+it}\right) }^{2}}+{\rho
^{2}}}\mp iz}}}\,,
\end{equation}%
where $\ a>0,$ which they derived by a Fourier synthesis and limiting the 
integration in $k$-space to one of the hemispheres (with ${{k}_{z}}>0$ or
${{k}_{z}}<0$, respectively). Here, we carry out the synthesis in a slightly
modified version that incorporates complex values ${{z}_{*}}=z+i{{z}_{s}}$,
${{z}_{s}}\equiv {({{a}_{2}}-{{a}_{1}})}/{2}\;$ of the $z$-coordinate and, thus,
a two-parameter solution like in Eq. (7). Since the solution is axially symmetric, 
we base the synthesis on the zeroth-order Bessel beams; specifically, 
\begin{widetext}
\begin{equation}
{h_{\pm }}\left( {\rho ,z,t}\right) ={\frac{1}{2}}\int_{-\infty }^{\infty }{d%
{k_{z}}}H\left( {\pm {k_{z}}}\right) {e^{i{k_{z}}z}}\int_{0}^{\infty }{%
d\kappa \,\kappa }{J_{0}}\left( {\kappa \rho }\right) {e^{-ict\;\sqrt{{%
\kappa ^{2}}+k_{z}^{2}}}}G\left( {\kappa ,{k_{z}}}\right) ,
\end{equation}
\end{widetext}
where $H\left( \cdot  \right)$ designates the Heaviside unit step function 
which ensures the integration in one hemisphere only. 
Choosing the spectrum 
\begin{equation}
G\left( {\kappa ,{k_{z}}}\right) ={\frac{1}{\sqrt{{\kappa ^{2}}+k_{z}^{2}}}}{%
e^{-ac\sqrt{{\kappa ^{2}}+k_{z}^{2}}}},\;\,a>0
\end{equation}%
and introducing the new variable $\lambda =\sqrt{{\kappa ^{2}}+k_{z}^{2}}$,
one obtains 
\begin{multline}
{h_{\pm }}\left( {\rho ,z,t}\right)=\\ {\frac{1}{2}}\int_{-\infty }^{\infty }{d%
{k_{z}}}H\left( {\pm {k_{z}}}\right) {e^{i{k_{z}}z}}\int_{{k_{z}}}^{\infty }{%
d\lambda }{J_{0}}\left( {\rho \sqrt{{\lambda ^{2}}-k_{z}^{2}}}\right) {%
e^{-a\lambda }}.
\end{multline}%
The integration over $\lambda $ is carried out \cite{LW29}, p.~191, No.~9, yielding 
\begin{multline}
{h_{\pm }}\left( {\rho ,z,t}\right) ={\frac{1}{{2\sqrt{{c^{2}}{{(a+ict)}^{2}}%
+{\rho ^{2}}}}}}\times\\\times\int_{-\infty }^{\infty }{d{k_{z}}}
H\left( {\pm {k_{z}}}%
\right) {e^{i{k_{z}}z-\left\vert {{k_{z}}}\right\vert \sqrt{{c^{2}}{{(a+ict)}%
^{2}}+{\rho ^{2}}}}}.
\end{multline}%
Finally, the integration over ${{k}_{z}}$ results in the unidirectional solutions
given in Eq.~(8), where $z$ is replaced by ${{z}_{*}},$ while the restriction
$\left| \operatorname{Im}{{z}_{*}} \right|<a$ avoids singularity. Note that 
the wavefunctions ${{h}_{\pm }}$ given in Eq.~(8), but modified in this manner,
are equivalent, respectively, to ${{{u}_{+}}}/{2}\;$ and ${{{u}_{-}}}/{2}\;$ 
in Eq.~(7), provided that, in addition, $a=\left( {{a}_{1}}+{{a}_{2}} \right)/2$. 

Without detailed discussion of its
unidirectional features, a seemingly different type of solution was derived
by Wong and Kaminer in 2017 \cite{LW30}; specifically, 
\begin{multline}
\psi \left( {\rho ,z,t}\right) =-i{\frac{{a+ict}}{{{k_{0}}{{\tilde{R}}%
^{2}}}}}\left( {{\frac{1}{{{k_{0}}\tilde{R}}}}{f^{-s-1}}+{\frac{{s+1}}{s}}{%
f^{-s-2}}}\right) ;\\
f=1-{k_{0}}\left( {iz+a-\tilde{R}}\right) /s,\; \tilde{R}=\sqrt{{{%
\left( {a+ict}\right) }^{2}}+{\rho ^{2}}},
\end{multline}
where $k_{0}=\omega _{0}/c=2\pi /\lambda _{0}$.
This solution can be derived from the Fourier synthesis in Eq. (1).
Assuming, first, the spectrum ${F_{1}}\left( {{k_{z}},k}\right) ={F_{2}}%
\left( {{k_{z}}}\right) \exp \left( {-ak}\right) ,$ we obtain 
\begin{equation}
\psi _{+}^{1}\left( {\rho ,z,t}\right) =\,{\frac{1}{{\tilde{R}}}}%
\int_{0}^{\infty }{d{k_{z}}}{e^{i{k_{z}}z}}{e^{-{k_{z}}\tilde{R}}}{F_{2}}%
\left( {{k_{z}}}\right) .
\end{equation}%
Let, next, ${k_{z}}/{k_{0}}=\chi /s$ and choose the spectrum so that 
\begin{equation}
\begin{split}
\psi _{+}^{1}\left( {\rho ,z,t}\right) &=\,{\frac{1}{{\tilde{R}}}}%
\int_{0}^{\infty }{d\chi }\,{e^{-\chi q}}{\frac{{{\chi ^{s}}}}{{\Gamma
\left( {s+1}\right) }}};\\q&=1-{\frac{{{k_{0}}}}{s}}\left( {iz+a-\tilde{R}}%
\right) .
\end{split}
\end{equation}%
The final solution assumes the form 
\begin{equation}
\psi _{+}^{1}\left( {\rho ,z,t}\right) =\,{\frac{1}{{\tilde{R}}}}{\left[ {1-{%
k_{0}}\left( {iz+a-\tilde{R}}\right) }\right] ^{-s-1}}={\frac{1}{{\tilde{R}}}%
}{f^{-s-1}}.
\end{equation}%
The solution given in Eq. (13) results from differentiation of 
$\psi _{+}^{1}\left( {\rho ,z,t}\right) /{k_{0}}$
with respect to time. 

All the unidirectional wavepackets discussed in this
section are finite-energy solutions of the three-dimensional scalar wave
equation in free space. An important question is how do they propagate in
the positive $z-$direction? We shall answer this question in connection to
the Bialynicki-Birula \textit{et al.} \cite{IBB} wavepacket which seems to
be the simplest. Consider the part of the solution involving $z$ and $t$;
specifically, 
\begin{equation}
z=-i\sqrt{{c^{2}}{{\left( {a+it}\right) }^{2}}+{\rho ^{2}}}.
\end{equation}%
Then, the real group speed is given by 
\begin{equation}
{{v}_{g}}\left( \rho ,t \right)=\operatorname{Re}\left\{ \frac{\partial }{\partial t}
z\left( \rho ,t \right) \right\}=c\operatorname{Re}\left\{ \frac{{{c}}
\left( a+it \right)}{\sqrt{{{c}^{2}}{{\left( a+it \right)}^{2}}+{{\rho }^{2}}}} \right\}.
\end{equation}
It is seen that the group speed depends both on the radial distance and time. 
A plot of the group speed (normalized with respect to the speed of
light in vacuum equal to unity) is shown in Fig. 1  for three values of $\rho .$
\begin{figure}[htb]
\centering
\includegraphics[width=8.5cm]{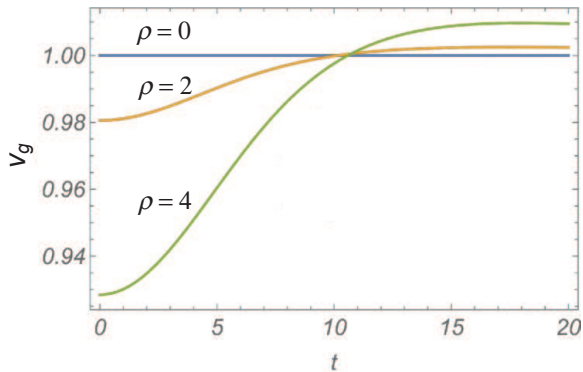}\caption{Plot of the normalized ($c=1$) 
group speed vs. time for $\rho =
0,\,\,2,\,\,4$ and parameter value $a = 10.$ Here and in the subsequent figures 
the parameters, coordinates, and time are given in dimensionless length units, 
see also the paragraph after Eq. (26).}
\end{figure}
On axis $\left(\rho =0\right) ,\;v_{g}=1$ for all values of time. At $%
\rho =2$, the speed is subluminal (but very close to unity) for small values
of time, it becomes luminal at a value of time slightly larger than 10,
superluminal afterwards, and tends to unity for very large values of time. A
similar behavior is exhibited for larger values of $\rho .$

This behavior is different from a finite-energy unidirectional scalar
wavepacket moving at a fixed speed. An example is provided in Appendix A for
a solution to the equation of acoustic pressure under conditions of uniform
flow.

\section{Extended Scalar Unidirectional Localized Waves}

 Courant and Hilbert \cite{LW31} have pointed out that a 
 "relatively undistorted" progressive solution to the homogeneous 
 three-dimensional (3D) scalar wave equation in vacuum assumes the form 
\begin{equation}
\psi (\vec{r},t)=\frac{1}{g(\vec{r},t)}\,f[\theta (\vec{r},t)],  \label{14}
\end{equation}
where $f(\cdot )$ is essentially an arbitrary function, $\theta (\vec{r},t)$%
, referred to as the "phase" function, is a solution to the nonlinear 
characteristic equation 
\begin{equation}
{\left( \frac{{\partial \theta }}{{\partial x}}\right) ^{2}}+{\left( \frac{{%
\partial \theta }}{{\partial y}}\right) ^{2}}+{\left( \frac{{\partial \theta 
}}{{\partial z}}\right) ^{2}}-\frac{1}{{c^{2}}}{\left( \frac{{\partial
\theta }}{{\partial t}}\right) ^{2}}=0,  \label{15}
\end{equation}
and $g(\vec{r},t)$ is an "attenuation" function; 
the latter depends on the choice of $\theta (\vec{r},t)$, but not
in a unique manner. Along this vein, a very general class of solutions to
the homogeneous scalar wave equation in free space is given as 
\begin{equation}
\begin{split}
\psi_{+}(\vec r,t)&=\frac{1}{g(\vec r,t)}f\left[
\theta(\alpha,\beta)\right]  ;\\ g(\vec r,t) &\equiv\sqrt
{\rho^{2}-c^{2}(t-it_{s})^{2}},\\ \alpha(\vec r,t) &\equiv
\sqrt{\rho^{2}-c^{2}(t-it_{s})^{2}}-i(z+iz_{s}),\\ \beta
(\vec r,t) &\equiv\frac{\rho \, e^{-i\phi}}{ic(t-it_{s}%
)+g(\vec r,t)},
\end{split}
\end{equation}
in polar coordinates. Here, ${z_{s}}$ and ${t_{s}}$ are free positive parameters.
In the sequel, we shall discuss in detail the specific
azimuthally asymmetric solution 
\begin{equation}
   {{\psi }_{+}}\left( \rho ,\phi ,z,t \right)=\ \frac{1}{2\,g(\vec{r},t)}
   \frac{1}{{{\alpha }^{q}}\left( \vec{r},t \right)}{{\,\text{e}}^{-p\alpha 
   \left( \vec{r},t \right)}}{{\beta }^{m}}\left( \vec{r},t \right), 
\end{equation}
 where $p$ is a positive free parameter. It should be noted that for
 $m=0,\,p=0$ and $q=1,$ the solution ${\psi }_{+}\left( \vec{r},t \right)$ 
 is identical to the Bialynicki-Birula azimuthally symmetric expression
 ${{h}_{+}}$ given in Eq. (8) if ${{z}_{s}}=0$ and ${{t}_{s}}=a$. Also, 
 for $m=0,\ p=0$ and $q=s+1$, ${{\psi }_{+}}\left( \vec{r},t \right)$ is
 a slight variation of the expression $\psi _{+}^{1}\left( \rho ,z,t \right)$
 in Eq. (16).

\section{Energy Backflow in Unidirectional Spatiotemporal Localized Waves}
\subsection{Scalar-valued wave theory}
The energy transport equation corresponding to the (3+1)D homogeneous scalar
wave equation 
\begin{equation}
	\left( {{\nabla }^{2}}-\frac{1}{{{c}^{2}}}\frac{{{\partial }^{2}}}{\partial
	{{t}^{2}}} \right)\psi \left( \vec{r},t \right)=0
	\end{equation}
governing the \textit{real-valued} wave function $\psi \left( \vec{r},t \right)$ 
in free space is given as \cite{49,50}
\begin{equation}
	\nabla \cdot \vec{S}+\frac{\partial }{\partial t}U=0,
	\end{equation}
where
	\begin{equation}U=\frac{1}{2}\frac{1}{{{c}^{2}}}{{\left( \frac{\partial }
{\partial t}\psi  \right)}^{2}}+\frac{1}{2}\nabla \psi \cdot \nabla \psi 
	\end{equation}
is the energy density $\left( J/{{m}^{3}} \right)$ and 
\begin{equation}
\vec{S}=-\frac{\partial }{\partial t}\psi \,\nabla \psi 
\end{equation}
is the energy flow density vector $\left( W/{{m}^{2}} \right)$.
 \begin{figure}[htb]
\centering
\includegraphics[width=8.5cm]{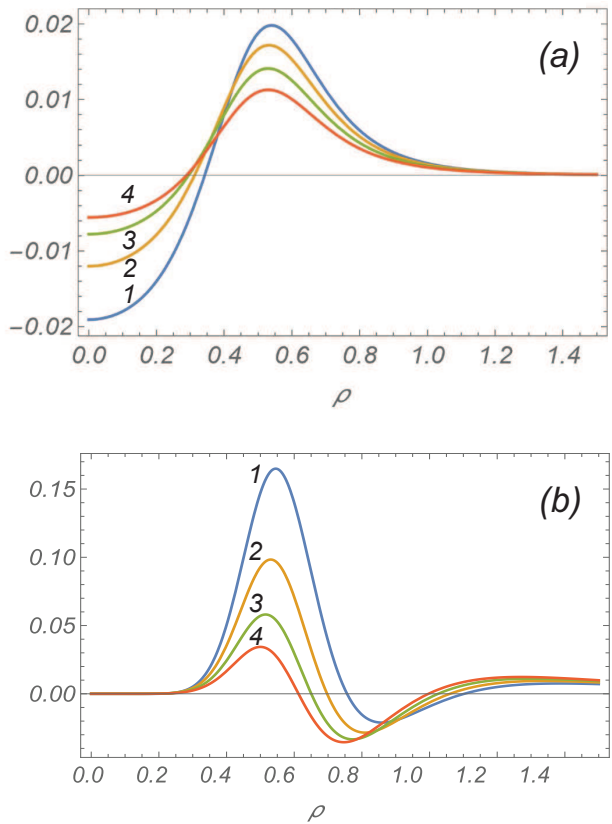}\caption{Plots of the longitudinal 
component of the energy flow vector ${{S}_{z}}$ at $t=0.5,$ with the speed of 
light normalized to unity. (a) $m=0,\, q=1,\,  p=0.$ (b) $m=3,\ q=1,\  p=0.$.
For curves 1, 2, 3, 4, values of the axial coordinate are z=0.9, z=1.0, z=1.1, and 
z=1.2, respectively. The plots are normalized with respect to their peak intensities at $t=0.$}
\end{figure}

An examination of energy backflow is accomplished by examining the properties of
the $z-$component of the energy flow vector corresponding to the real part of 
the extended unidirectional scalar complex wave function ${{\psi }_{+}}\left( \rho
,\phi ,z,t \right)$ given in Eq.~(22). Plots are shown in Fig.~2 of ${{S}_{z}}$ 
versus $\rho $ for four axial positions, first for $m=0,\ p=0,\ q=1$ and 
then, in Fig.~2(b) for $m=3,\ p=0,\ q=1.$  For both plots the parameter values are
${{t}_{s}}=0.3\text{ and}\ {{z}_{s}}=0.1$. The time is fixed at $t=0.5$, with the 
speed of light of vacuum normalized to unity. Thus, the parameters and time are 
given in dimensionless length units and therefore the results are applicable not 
only in optical but also in microwave, etc. regions.  For example, if the chosen 
values of the parameters $t_{s}$ and $z_{s}$, as well as the time $t$ were in
micrometers, then $t=0.5$ would correspond to 1.67 femtoseconds which would also be 
approximately equal to the  pulsewidth.  
 The value $t=0.5$ has been chosen since at $t=0$ the backflow is absent but at 
 higher values of $t$ the pulse would spread out too much and the energy flow would 
 become negligible.
 
We studied the simplest case with $m=0,\ p=0,\ q=1$ in more detail by plotting the 
local energy transport velocity vector field, $\vec{V}=\vec{S}/U$, see Fig.~3.
 \begin{figure}[htb]
\centering
\includegraphics[width=8.5 cm]{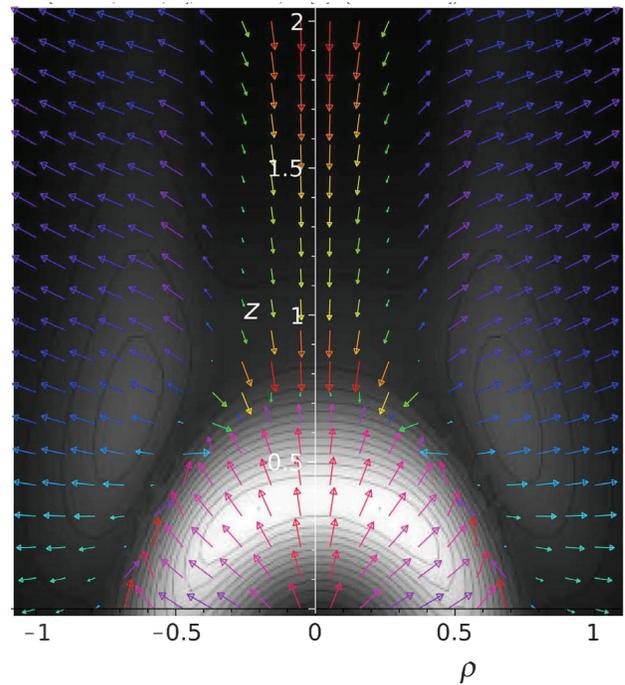}\caption{Energy flow velocity vector field 
$\vec{V}(\rho,z)$ as the ratio of the energy flow vector $\vec{S}$ and energy 
density $U$, plotted at 20x20 spatial points in the plane $(x,z)$, 
where $x=\pm\rho$ represents any transverse axis. 
The background image is a grey-scale plot (with contour lines) of  $\sqrt{U}$ and shows 
the pulse and its two weaker side-maxima at the instant $t=0.5$ of their stage of evolution. 
The maximum length of arrows corresponds to $c$ (or to 1 of the dimensionless velocity).
In addition to the orientation of the arrows, values of the projection $V_{z}$ are 
expressed by colors of the arrows in the on-line version: the color gamut 
violet-indigo-blue-cyan-green-yellow-red
corresponds to range from +1  to -1 (cyan corresponds to 0).
Other parameters are the same as in Fig.~2(a). }
\end{figure}
The plots demonstrate that at the instant $t=0.5$ energy flows backward in the region
$(\rho<0.3,z>0.5)$. In a sense the situation resembles the sea drawback phenomenon, 
where the water recedes ahead of the tsunami wave peak.

A complex-valued version of the wavefunction of Eq.~(22) exhibits practically no 
backflow effect because for its imaginary part the backflow regions have different 
locations as compared to those of the real part. Therefore, negative values of $S_z$ 
from the real part are compensated by much stronger positive values of $S_z$ from 
the imaginary part and \textit{vice versa} since the energy flow vectors from both 
parts sum up additively. The same holds for the wavefunction of Eq.~(5) or Eq.~(4).

An interesting question is whether the power flux, that is the integral of 
the $z-$component of the energy flow vector over a transverse plane at fixed 
values of $z$ and $t$ can be negative, at least  for a small circular disk. 
Indeed, this is the case. From Fig. 2(a), we use the dimensionless values 
$z=0.9$ and $t=0.5.$ Then, the integration results in the plot are shown in Fig.~4.
\begin{figure}[htb]
\centering
\includegraphics[width=8.5cm]{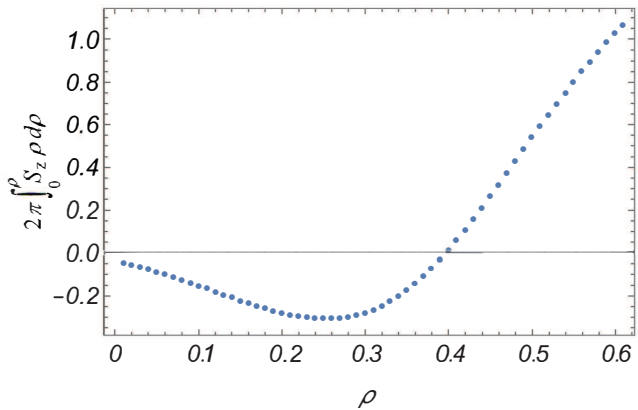}\caption{Plot of the numerical evaluation of the power through a circular disk, with values of the radial coordinate in the range $\left\{ 0,\ 0.6 \right\}$ in intervals of 1/100  at the 
fixed values $z=0.9$  and $t=0.5$, with the speed of light normalized to unity. The plot corresponds to that in Fig.~2(a) for $m=0$, $p=0$, $q=1$ and the parameter values ${{t}_{s}}=0.3\text{ and}\ {{z}_{s}}=0.1$.}
\end{figure}

We see that the negative flux increases in absolute value up to the radius 
$\simeq0.26$ of the disk, in full accordance with Fig.~3.

To conclude, it is somewhat surprising that a scalar solution to the wave equation
exhibits the backflow effect because earlier studies have instilled an opinion that
the effect appears in electromagnetic fields of specific polarization. Our results
indicate that backflow is possible not only in optical fields describable by scalar
approximation but also in acoustical fields. 

\subsection{Vector-valued wave theory}
Implicit in the article by Bialynicki-Birula \textit{et al.} \cite{IBB} is that 
the examination of the energy backflow characteristics of a vector-valued 
unidirectional localized wave is based on a complex Riemann-Silberstein vector 
\cite{51,52} derived from the vector Hertz potential 
$\vec{\Pi }=2\left( {{{\vec{a}}}_{x}}+i{{{\vec{a}}}_{y}} \right)\psi $
in Cartesian coordinates, or 
$\vec{\Pi }=\exp \left( i\phi  \right)\left( {{{\vec{a}}}_{\rho }}+i{{{
\vec{a}}}_{\phi }} \right)\psi$  
in cylindrical coordinates, where the complex-valued wave function 
$\psi \left( \vec{r},t \right)$
is a solution of the (3+1)D scalar wave equation in free space; specifically, 
the function ${{h}_{+}}$ in Eq. (8). The complex-valued Riemann-Silberstein 
vector is defined as 
\begin{equation}
\vec{F}=\nabla \times \nabla \times \vec{\Pi }+\frac{i}{c}\frac{\partial }{
\partial t}\nabla \times \vec{\Pi }.
\end{equation}
It obeys the equations
\begin{equation}
\nabla \times \vec{F}-\frac{i}{c}\frac{\partial }{\partial t}\vec{F}=0,
\quad \nabla \cdot \vec{F}=0,
\end{equation}
that are exactly equivalent to the homogeneous Maxwell equations for the 
free-space real electric and magnetic fields $\vec{E}\text{ and }\vec{B}$, 
defined in terms of $\vec{F}$ as follows:
\begin{equation}
\vec{F}=\sqrt{\frac{{{\varepsilon }_{0}}}{2}}\left( \vec{E}+ic\vec{B} \right).
\end{equation}
The importance of the specific choice for the vector Hertz potential is that the
corresponding Riemann-Silberstein vector, with the scalar wave function 
$\psi \left( \vec{r},t \right)$ being \textit{any luminal spatiotemporally 
localized wave}  [e.g., the splash mode in Eq. (6)], is null, that is, it has 
the property $\vec{F}\cdot \vec{F}=0$. Equivalently, the two Lorentz-invariant
quantities ${{I}_{1}}=\vec{E}\cdot \vec{B}$ and ${{I}_{2}}={{\left| {\vec{E}}
\right|}^{2}}-{{c}^{2}}{{\left| {\vec{B}} \right|}^{2}}$ are both equal to zero.
Under certain restrictions, the resultant field can be a pure Hopfion \cite{53}
exhibiting interesting topological properties, such as linked and knotted field
lines, or a Hopfion-like structure, such as the one established by Bialynicki-Birula
\textit{et al.} \cite{IBB}.

In the discussion below, the formalism described above will be followed, however  based on the simpler vector Hertz potential  
$\vec{\Pi }=\psi {{\vec{a}}_{z}}.$ 
The Poynting vector, defined in terms of the real fields as $\vec{S}=\vec{E}\times 
\vec{H},$ can be written in terms of the Riemann-Silberstein vector and its complex 
conjugate as follows: $\vec{S}=-i{{\vec{F}}^{*}}\times \vec{F}.$ An examination of 
energy backflow will be accomplished by examining the properties of the $z-$component 
of the Poynting vector corresponding to the extended unidirectional scalar complex 
wave function ${{\psi }_{+}}\left( \rho ,\phi ,z,t \right)$ given in Eq. (22). 
 \begin{figure}[h]
\centering
\includegraphics[width=8.5cm]{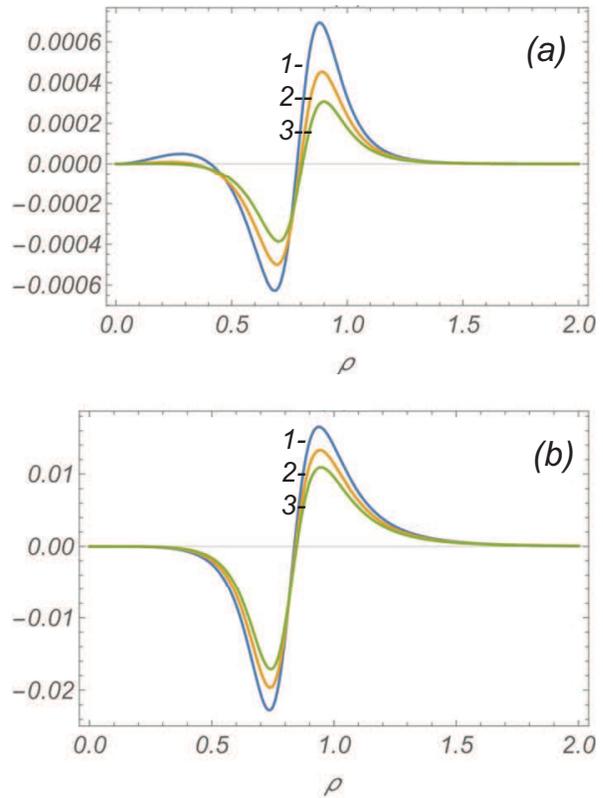}\caption{Plots of the longitudinal component
of the Poynting vector ${{S}_{z}}$ at $t=0.8,$ with the speed of light normalized to
unity. (a) $m=0,\ p=0,q=1.$ (b) $m=3,p=0,\ q=1.$ For curves 1, 2, 3, values of the 
axial coordinate are $z=1.2$, $z=1.3$, and $z=1.4$ respectively.  
For both plots the parameter values
are ${{t}_{s}}=0.2\text{ and }{{z}_{s}}=0.1.$ Both plots are normalized with respect
to their peak intensities at $t=0.$ }
\end{figure}
Plots are shown in Fig.~5 of  ${{S}_{z}}$ versus $\rho $ for 
three axial positions, first for $m=0,\ p=0,q=1$ and then for $m=3,\ p=0,q=1.$ 
Note that the values of $t$ and $z$, as well as of the parameters ${{t}_{s}}
\text{ and }{{z}_{s}}$ differ from those of Fig.~2. We see that the vector-valued versions of both waves in a certain spatiotemporal 
region exhibit the backflow effect which is weak but comparable to that of the scalar-valued waves.

\section{Unidirectional Hopfion-like spatiotemporally localized wave without energy backflow}
The basic (or pure) Hopfion \cite{53} is a finite-energy luminal spatiotemporally 
localized solution to Maxwell’s equations in free space with unique topological properties. 
Specifically, all electric and magnetic field lines are closed loops, and any two electric (or magnetic) field lines are linked once with one another. However, the basic Hopfion has equally distributed forward (along the positive $z-$direction) and backward components. 
By construction, then, it exhibits energy backflow. The unidirectional Hopfion-like 
wave structure in \cite{IBB} shares some of the topological characteristics with the 
pure Hopfion but exhibits energy backflow.
\begin{figure}[h]
\centering
\includegraphics[width=5 cm]{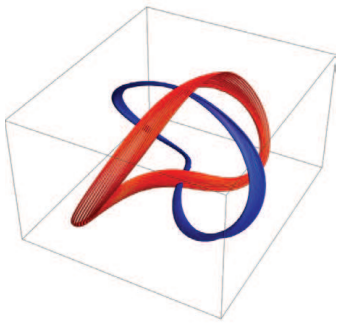}\caption{Linkages of electric and magnetic field 
lines at a fixed value of time, with the speed of light in vacuum normalized to unity.  
The parameter values are ${{t}_{s}}=0.3$ and ${{z}_{s}}=0.1.$ }
\end{figure}
A unidirectional Hopfion-like wave packet devoid of energy backflow will be constructed 
in this section. Toward this goal, a technique due originally to Whittaker \cite{54} and 
Bateman \cite{55} (see, also, \cite{56,57} for modern applications) will be used. First, 
the following two quantities, $\overline{\alpha } \left( \vec{r},t \right)$ and 
$\overline{\beta} \left( \vec{r},t \right)$, known as
\textit{Bateman conjugate functions}, will be defined in terms of the functions
$\alpha \left( \vec{r},t \right)$ and $\beta \left( 
\vec{r},t \right)$ in Eq. (21):
\begin{equation}
\bar{\alpha }\left( \vec{r},t \right)=\frac{1}{{{\alpha }^{*2}}\left( \vec{r},t \right)},
\ \bar{\beta }\left( \vec{r},t \right)={{\beta }^{*}}\left( \vec{r},t \right).
\end{equation}
Any functional of these two functions obeys the nonlinear characteristic Eq. (20). 
Furthermore, these two functions obey the \textit{Bateman constraint }
\begin{equation}
\nabla \bar{\alpha }\times \nabla \bar{\beta }-\frac{i}{c}\left( \frac{\partial \bar{\alpha }}
{\partial t}\nabla \bar{\beta }-\frac{\partial \bar{\beta }}{\partial t}\nabla \bar{\alpha } \right)=0.
\end{equation}
Under these assumptions, the complex vector 
\begin{equation}
\vec{F}=\nabla \bar{\alpha }\times \nabla \bar{\beta } 
\end{equation}
is a \textit{null Riemann-Silberstein vector} governed by the expressions in Eq. (28)
and related to the real electric and magnetic fields as given in Eq. (29). These fields 
have some of the topological characteristics of those associated with a pure Hopfion. 
Fig.~6 shows the linkages of the electric and magnetic field lines. Similar 
linkages characterize any two electric (or magnetic) field lines. The basic Hopfion is 
characterized by linked single closed field line loops. In our case, we have 
linked bundles  of field lines instead.
 
The $z-$component of the Poynting vector $\vec{S}=-i{{\vec{F}}^{*}}\times \vec{F}$ 
is plotted in Fig.~7. No energy backflow is present in this case. 
\begin{widetext}
 \begin{figure}[htb]
\centering
\includegraphics[width=17cm]{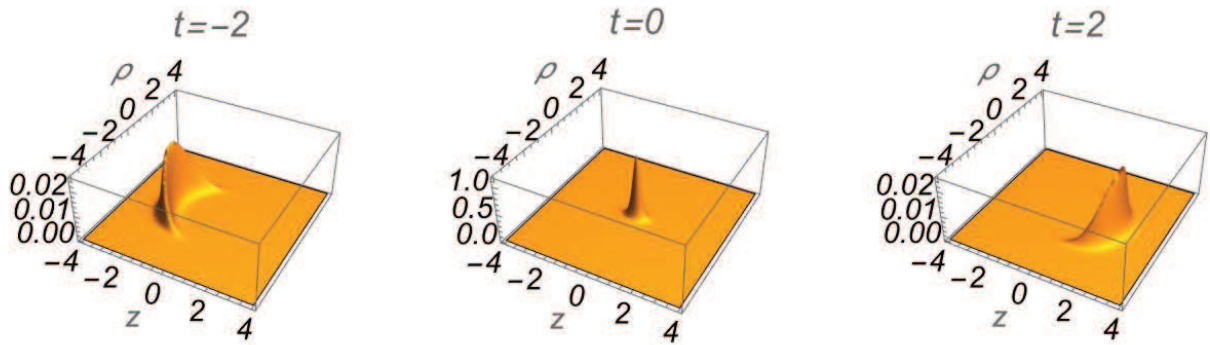}\caption{Plot of the $z-$component of the Poynting 
vector versus $z$ and $\rho $ for three values of time, $t$.  The parameter values are 
${{t}_{s}}=0.3$ and ${{z}_{s}}=0.1.$}
\end{figure}
\end{widetext}
The ratio of the Poynting vector and the electromagnetic volume density is the 
local energy transport velocity  $\vec{V}\left( \vec{r},t \right)=\vec{S}/U$, 
with the energy density given in terms of the
Riemann-Silberstein vector as $U\left( \vec{r},t \right)={{\vec{F}}^{*}}\cdot \vec{F}$.
Due to the nullity of $\vec{F},$ the modulus of the energy transport velocity is 
equal to the speed of light $c,$ although $\vec{V}\left( \vec{r},t \right)$
may vary in space and time. In the case of the basic Hopfion, the local energy 
velocity depends on the $z\text{ and }t$ through the combination $z-ct;$ that is, 
it evolves along the $z$- direction without any deformation. Such a structure is 
known as a Robinson congruence. In the case under consideration in this section 
the local energy transport velocity is altogether independent of the 
coordinate $z.$ 
The plot in Fig.~8 shows the $z-$component of the local energy 
transport velocity versus $\rho $ for three values of time. 
 \begin{figure}[h]
\centering
\includegraphics[width=7 cm]{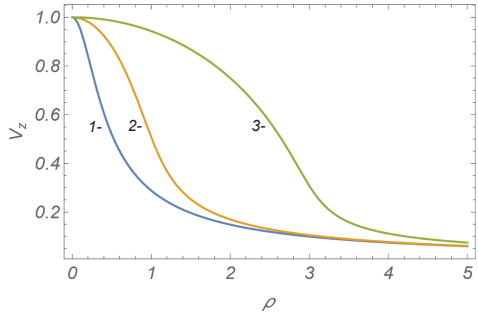}\caption{Plot of the $z-$component of the 
local energy transport velocity versus $\rho $ for three values of time, $t=0.1,\ 1$ 
and $t=3.$}
\end{figure}
The absence of energy backflow is clearly evident.
\section{Concluding remarks}
As already has been mentioned, it is crucial whether the plane-wave constituents of 
a localized wavepacket propagate only in the positive z-direction or, also, backward.
In the first case, not only the wave can be launched from an aperture as a freely
propagating beam but, also, the very question of energy backflow is meaningful. 
Since most of the analytically constructed finite-energy spatiotemporally localized
waves (luminal, subluminal, and superluminal) are acausal in the sense that they
include both forward and backward propagating components, several attempts have been
made to create close replicas of such waves that can be causally launched as forward
beams from apertures (see, e.g., \cite{LW6} and \cite{58}), or even derive exact causal
unidirectional wavepackets \cite{59,60}. Based usually on the Huygens principle, 
dealing with the former is computationally intensive. On the other hand, the latter
are quite complicated analytically. The study of energy backflow in this article 
has been confined to relatively simple causal unidirectional finite-energy localized
waves arising from a factorization of the basic splash mode \cite{LW3,LW4}. 
Specific results are given for the energy backflow exhibited in known azimuthally 
symmetric unidirectional wavepackets, as well as in novel azimuthally asymmetric
extensions. Using the Bateman-Whittaker technique, a novel finite-energy
unidirectional null localized wave has been constructed that is devoid of 
energy backflow and has some of the topological properties of the basic Hopfion.

The study of energy backflow of the vector-valued unidirectional localized study in 
Sec. 4 was based on the Riemann-Silberstein complex vector that results in electric 
and magnetic fields that both have nonzero $z$ components (non-TE and non-TM). 
Although specific results have not been incorporated in this article, an examination 
of the Poynting vectors associated individually with pure TE and TM fields associated 
to the scalar unidirectional localized wave ${{\psi }_{+}}\left( \rho ,\phi ,z,t \right)$ 
given in Eq.~(22) shows the presence of energy backflow. This is altogether different 
from the cases of the superposition of four plane waves \cite{Katz, quartet,meiequartet}, 
Bessel beams \cite{AriBBnegS, TM+TE}, and a pulsed electromagnetic X wave \cite{TM+TExwave}, 
all of which require a superposition of TE and TM fields for the appearance of energy backflow.

Finally, some remarks about possible experimental studies and relation of
our results to quantum optics.

As was said already in Section III, the results (incl. numerical plots) are
applicable irrespectively of the frequency range of the EM field. At low
frequencies up to the microwave region, the measurements for studying the
non-stationary behavior of Poynting vector in pulsed fields considered here
require time resolution up to nanosecond range. For that purpose, known
sensor-based techniques developed for monochromatic fields (see, e. g., \cite
{61}) would be applicable. However, to the best of our knowledge, no
experiments have been carried out for electromagnetic or acoustic energy
backflow associated with pulsed spatiotemporally localized wavefunctions.

As far as experiments on the electromagnetic energy backflow effect in the
optical range is concerned, to date a few studies with monochromatic fields
can be found: in addition to Refs. 11 and 12, studies of the effect in
nanoscale focuses \cite{62,63} are appropriate to mention here.
Distribution of the Poynting vector in optical fields can be measured via
the motion of probe Rayleigh particles or via investigation of polarization
in passage through an anisotropic crystal \cite{64}. However, optical
experiments on the fields considered in our paper are hardly feasible today
because they would need near-single-cycle light pulses
and---consequently---sub-femtosecond temporal resolution.

The expressions derived in this paper also apply to quantum optics: as it is
known, the spatio-temporal dependence of the (quantum mechanical) wave
function of a single photon, treated as a particle-like object, is given by
the Riemann--Silberstein vector of the corresponding classical EM field.
Moreover, as shown in \cite{64}, in terms of quantum weak measurements,
observation of averaged trajectories of single photons can be considered as
measurement of the distribution of the Poynting vector in the corresponding
classical optical field, incl. backflow effects.

\section{Appendix A}

The equation of acoustic pressure under conditions of uniform flow is given
as follows: 
\begin{equation}
\left[ {{\nabla ^{2}}-{\frac{1}{{u_{0}^{2}}}}{{\left( {{\frac{\partial }{{%
\partial t}}}+\vec{u}\cdot \nabla }\right) }^{2}}}\right] p\left( {\vec{r},t}%
\right) =0.  \tag{A1}\label{A1}
\end{equation}%
Here, ${u_{0}}$ is the speed of sound in the rest frame of the medium and $%
\vec{u}$ is the uniform velocity of the background flow. In the special case
where $\vec{u}=u{\vec{a}_{z}}$ and $u={u_{0}}$ the resulting equation for the
acoustic pressure simplifies as follows: 
\begin{equation}
\left( {\nabla _{t}^{2}-{\frac{2}{{{u_{0}}}}}{\frac{{{\partial ^{2}}}}{{%
\partial t\partial z}}}-{\frac{1}{{u_{0}^{2}}}}{\frac{{{\partial ^{2}}}}{{%
\partial {t^{2}}}}}}\right) p\left( {\vec{r},t}\right) =0.  \tag{A2}\label{A2}
\end{equation}%
Under this assumption, several exact analytical infinite-energy
nondiffracting and finite-energy slowly nondiffracting spatiotemporally 
localized wave solutions are supported.
One such solution is the finite-energy unidirectional splash-like  mode 
\begin{multline}
p\left( {\rho ,z,t}\right) =\\={\frac{1}{{{a_{1}}+i\left( {z-2{u_{0}}t}\right) }%
}}{\left( {{a_{2}}-iz+{\frac{{{\rho ^{2}}}}{{{a_{1}}+i\left( {z-2{u_{0}}t}%
\right) }}}}\right) ^{-q}},  \tag{A3}\label{A3}
\end{multline}%
with ${a_{1,2}}$ positive parameters. This is a finite-energy unidirectional
wavepacket moving along the $z-$direction at the fixed speed $v=2{u_{0}}.$


\begin{thebibliography}{99}

\bibitem{Berry} M. V. Berry, Quantum backflow, negative kinetic energy, and
optical retro-propagation, J. Phys. A: Math. Theor. \textbf{43}, 415302
(2010).

\bibitem{IBB} I. Bialynicki-Birula, Z, Bialynicki-Birula and S.
Augustynowicz, Backflow in relativistic wave equations, J. Phys. Math. Gen. 
\textbf{55}, 255702 (2022)

\bibitem{Bracken} A J. Bracken, Probability flow for a free particle: new
quantum effects, Phys. Scr. \textbf{96}, 045201 (2021).

\bibitem{Comment} M. Barbier, C. J. Fewster, A. Goussev, G. Morozov, S. C.
L. Srivastava, Comment on "Backflow in relativistic wave equations",
arXiv:2210.05368 [quant-ph] 11 October 2022.

\bibitem{Katz} B. Z. Katsenelenbaum, What is the direction of the Poynting
vector? J. Commun., Technol. Electron., \textbf{42,} 119 (1997).

\bibitem{quartet} X.-L. You, C.-F. Li, Dependence of Poynting vector on
state of polarization, arXiv:2009.04119 [physics.optics] 5 Jan 2021.

\bibitem{meiequartet} P. Saari and I. Besieris, Backward energy flow in
simple four-wave electromagnetic fields, Eur. J. Phys. \textbf{42}, 055301
(2021).

\bibitem{negSfocal} B. Richards and E. Wolf, Electromagnetic diffraction in
optical systems, II. Structure of the image field in an aplanatic system,
Proc. R. Soc\textit{.} A \textbf{253 }358 (1959).

\bibitem{negSfocaluus} V. V. Kotlyar, S. S. Stafeev, A G. Nalimov, A. A.
Kovalev, and A. P. Porfirev, Mechanism of formation of an inverse energy flow in
a sharp focus, Phys. Rev. A, \textbf{101,} 033811 (2020).

\bibitem{negSfocaluus2} H. Li, C. Wang, M. Tang, and A. Xinzhong, Controlled
negative energy flow in the focus of a radial polarized optical beam, Opt.
Express \textbf{28} 18607 (2020).

\bibitem{BFexp1} Y. Eliezer, T. Zacharias, and A. Bahabad, Observation of
optical backflow, Optica \textbf{7}, 72 (2020).

\bibitem{BFexp2} A. Daniel,\ B.\ Ghosh, B.\ Gorzkowski, and R. Lapkiewicz,
Demonstrating backflow in classical two beams' interference,
arXiv:2206.05242 [physics.optics] 10 June 2022.

\bibitem{AriBBnegS} J. Turunen, A. T. Friberg, Self-imaging and
propagation-invariance in electromagnetic fields, Pure Appl. Opt. \textbf{2,}
51 (1993).

\bibitem{TM+TE} A. V. Novitsky, D. V. Novitsky, Negative propagation of
vector Bessel beams, J. Opt. Soc. Am\textit{. }A,\textit{\ }\textbf{24}%
\textit{\ }2844 (2007).

\bibitem{TM+TExwave} M. A. Salem and H. Ba\u{g}c\i , Energy flow
characteristics of vector X-Waves, Opt. Express\textit{\ }\textbf{19}, 8526
(2011).

\bibitem{LW1} J. N. Brittingham, Focus wave modes in homogeneous Maxwell
equations: Transverse electric mode, J. Appl. Phys. \textbf{54}, 1179-1189
(1983).

\bibitem{LW2} A. P. Kiselev Modulated Gaussian beams, Radio Phys. Quant.
Electron. \textbf{26}, 1014-1020 (1983).

\bibitem{LW3} R. W. Ziolkowski, Localized transmission of electromagnetic
energy, Phys. Rev. A \textbf{39}, 2005-2033 (1989).

\bibitem{LW4} I. M. Besieris, A. M. Shaarawi and R. W. Ziolkowski, A
bidirectional traveling plane wave representation of exact solutions of the
scalar wave equation, J. Math. Phys. \textbf{30}, 1254-1269 (1989).

\bibitem{LW5} J. Y. Lu and J. F. Greenleaf, Nondiffracting X waves-exact
solutions to the free space scalar wave equation and their finite aperture
realization, IEEE Trans. Ultrason. Ferroelectr. Freq. Contr. \textbf{39},
19-31 (1992).

\bibitem{LW6} R. W. Ziolkowski, I. M. Besieris and A. M. Shaarawi, Aperture
realizations of the exact solutions to homogeneous-wave equations, J. Opt.
Soc. Am. A \textbf{10}, 75-87 (1993).

\bibitem{LW7} P. Saari and K. Reivelt, Evidence of X-haped
propagation-invariant localized light waves, Phys. Rev. Lett. \textbf{79},
4135-4137 (1997).

\bibitem{LW8} I. M. Besieris, M. Abdel-Rahman, A. M. Shaaraw and A.
Chatzipetros, Two fundamental representations of localized pulse solutions
to the scalar wave equation, Progr. Electromagn. Res. (PIER) \textbf{19},
1-48 (1998).

\bibitem{LW9} J. Salo, J. Fagerholm, A. T. Friberg and M. M. Saloma, Unified
description of X and Y waves, Phys. Rev. E \textbf{62,} 4261 (2000).

\bibitem{LW10} R. Grunwald, V. Kebbel, U. Neumann, A. Kummrow, M. Rini, R.
T. Nibbering, M. Piche, G. Rousseau and M. Fortin, Generation and
characterization of spatially and temporally localized few-cycle optical
wave packets, Phys. Rev. A \textbf{67}, 063820-1-5 (2003).

\bibitem{LW11} P. Saari and K. Reivelt, Generation and classification of
localized waves by Lorentz transformations in Fourier space, Phys. Rev. E 
\textbf{69}, 036612-1-12 ( 2004).

\bibitem{LW12} S. Longhi, Spatial-temporal Gauss-Laguerre waves in dispersive
media, Phys. Rev. E \textbf{68}, 066612 1-6 (2003).

\bibitem{LW13} C. Conti, S. Trillo, P. di Trapani, G. Valiulis, A.
Piskarskas, O. Jedrkiewicz, and J. Trull, Nonlinear electromagnetic X waves,
Phys. Rev. Lett. \textbf{90}, 170406 1-4 (2003).

\bibitem{LW14} A. P. Kiselev, Localized light waves: Paraxial and exact
solutions of the wave equation (review), Opt. Spectrosc. \textbf{102},
603-622 (2007).

\bibitem{LW15} M. Yessemov, B. Bhaduri, H. E. Kondaksi and A. F. Abouraddy,
Classification of propagation-invariant space-time wavepackets in free spac:
Theory and experiments, Phys. Rev. A \textbf{99}, 023856 (2019).

\bibitem{LW16} \textit{Localized Waves}, edited by H. E. Hernandez-Figueroa,
M. Zamboni-Rached, and E. Recami (J. Wiley, New York, 2007).

\bibitem{LW17} \textit{Non-Diffracting Waves}, edited by H. E.
Hernandez-Figueroa, E. Recami, and M. Zamboni-Rached (J. Wiley, New York,
2013).

\bibitem{LW18} M. Yessemov, L. A. Hall, K. L. Schepler and A. F. Abouraddy, 
Space-time wavepackets, Advances in Optics and Photonics \textbf{14}, 455-570
 (2022).

\bibitem{LW19} K. Reivelt and P. Saari, Experimental demonstration of
realizability of optical focus wave modes, Phys. Rev. E \textbf{66},
056611-1-9 (2002).

\bibitem{LW20} P. Bowlan, H. Valtna-Lukner, M. L\~{o}hmus, P. Piksarv, P.
Saari, and R. Trebino, Measurement of the spatio-temporal field of
ultrashort Bessel-X pulses, Opt. Lett. \textbf{34}, 2276 (2009).

\bibitem{LW21} P. Saari, X-Type Waves in Ultrafast Optics, in \textit{\cite
{LW17}} pp. 109-134.

\bibitem{LW22} H. F. Kondakci and A. F. Abouraddy, Diffraction-free
space-time light sheets, Nat. Photon. \textbf{11}, 733-740 (2017).

\bibitem{LW23} B. Bhaduri, M. Yessenov, and A. F. Abouraddy, Space-time wave
packets that travel in optical materials at the speed of light in vacuum,
Optica \textbf{6}, 139-145 (2019).

\bibitem{LW24} N. Papasimakis, T. Raybould, V. A. Fedotov, D. P. Tsai, I.
Youngs and N. I. Zheludev, Pulse generation scheme for flying
electromagnetic doughnuts, Phys. Rev. B \textbf{9}7, 201409-1-6 (2018).

\bibitem{LW25} D. Comite, W. Fuscaldo, S. K. Podilchak, and V. G\'{o}%
mez-Guillam\'{o}n Buenndia, Microwave generation of X-waves by means of
planar leaky-wave antenna, Appl. Phys. Lett. \textbf{113}, 144102-1-5 (2018).

\bibitem{LW26} W. Fuscaldo, D. Comite, A. Boesso, P. Baccarelli, P.
Bughignoli and A. Galli, Focusing leaky waves: a class of electromagnetic
localized waves with complex spectra, Phys. Rev. Appl. \textbf{9},
054005-1-15 (2018).

\bibitem{MinuPRA2018} P. Saari, Reexamination of group velocities of
structured light pulses, Phys. Rev. A, \textbf{97}, 063824 (2018).

\bibitem{POY2019} P. Saari, O. Rebane, and I. Besieris, Energy-flow
velocities of nondiffracting localized waves, Phys. Rev. A \textbf{100},
013849 (2019).

\bibitem{LW27} J. Lekner, Electromagnetic pulses, localized and causal,
Proc. R. Soc. A \textbf{474}, 20170655 (2018).

\bibitem{LW29} \textit{Tables of integral transforms}, edited by A. Erdelyi
(McGraw-Hill, New York, 1954), Vol. I.

\bibitem{LW28} I. A. So, A. B. Plachenov and A. P. Kiselev, Simple
unidirectional finite-energy pulses, Phys. Rev. A \textbf{102}, 063529
(2020).

\bibitem{LW30} L. J. Wong and I. Kaminer, Abruptly focusing and defocusing
needles of light and closed-form electromagnetic wavepackets, ACS Photonics 
\textbf{4}, 1131 (2017).

\bibitem{LW31} R. Courant and D. Hilbert, \textit{Methods of Mathematical
Physics} (Interscience, New York, 1962), Vol. II.

\bibitem{49} H. S. Green and E. Wolf, A scalar representation of electromagnetic 
fields, Proc. Phys. Soc. A \textbf{66}, 1129 (1953).

\bibitem{50} L. Mandel and E. Wolf, \textit{Optical Coherence and Quantum Optics} 
(Cambridge University Press, Cambridge, UK, 1955), p. 288.

\bibitem{51} H. Weber, \textit{Die partiellen Differential-Gleichungen der 
mathematischen Physik nach Riemann’s Vorlesungen}, Friedrich Vieweg und Sohn, Brunschweig, 1901.

\bibitem{52} L. Silberstein, Electromagnetische Grundgleichungen in bivectorieller 
Behandlung, Ann D. Phys. \textbf{22}, 579-686 (1907).

\bibitem{53} A. Ranãda, Knotted solutions of the Maxwell equations in vacuum, 
J. Phys. A: Math. Gen.  \textbf{23}, L815-L820 (1990).

\bibitem{54}  E. T. Whittaker, On the expressions of the electromagnetic field due 
to electrons by means of two scalar potential functions, Proc. London Math. Soc. \textbf{1}, 367-372 (1904).

\bibitem{55}  H. Bateman, \textit{The Mathematical Analysis of Electrical and Optical 
Wave-Motion on the Basis of Maxwell’s Equations} (Dover, New York, 1955). 

\bibitem{56}  I. M. Besieris and A. M. Shaarawi, Hopf-Ranãda linked and knotted light 
beam solution viewed as a null electromagnetic field, Opt. Lett. \textbf{34}, 3887-3889 (2009).

\bibitem{57}  H. Kedia, I. Bialynicki-Birula, D. Peralta-Salas and W. T. M. Irvine, 
Tying knots in light fields, Phys. Rev. Lett. \textbf{111}, 150404 (2017).

\bibitem{58} A. M. Shaarawi, Comparison of two localized wave fields generated from dynamic 
apertures, J. Opt. Soc. Am. A \textbf{14}, 1804-1815 (1997).

\bibitem{59} C. J. R. Sheppard and P. Saari, Lommel pulses: An analytic form for localized 
waves of the focus wave mode type with bandlimited spectrum, Opt. Express \textbf{16}, 150-160 (2008).

\bibitem{60} M. Zamboni-Rached, Unidirectional decomposition method for obtaining exact 
localized solutions totally free of backward components, Phys. Rev. A \textbf{79}, 013816 (2009).
\bibitem{61} C. C. Chen and J. F. Whitaker, An optically-interrogated
microwave-Poynting vector sensor using cadmium manganese telluride. Opt.
Express \textbf{18}, 12239 (2010).

\bibitem{62} G. Yuan, E. F. Rogers and N. I. Zheludev, \textquotedblleft
Plasmonics\textquotedblright\ in free space: observation of giant
wavevectors, vortices, and energy backflow in superoscillatory optical
fields, Light: Sci. Appl. \textbf{8}, 1 (2019).

\bibitem{63} V. V. Kotlyar, S. S. Stafeef, A. G. Nalimov, A. A. Kovalev and
A. P. Porfirev, Experimental investigation of the energy backflow in the
tight focal spot, Comp. Opt. \textbf{44}, 863 (2020).

\bibitem{64} K. Y. Bliokh, A. Y. Bekshaev, A.G. Kofman, and F. Nori, Photon
trajectories, anomalous velocities and weak measurements: a classical
interpretation, New J. Phys. \textbf{15}, 073022 (2013).

\end{thebibliography}
\end{document}